\documentclass[iop,apjl,numberedappendix]{emulateapj}
\usepackage[latin1]{inputenc}
\usepackage{amssymb}
\usepackage{amsmath}
\usepackage{graphicx}
\usepackage{xspace}
\usepackage{natbib}
\usepackage{url}
\usepackage{paralist}
\usepackage{epsfig}

\def\cm{{\rm\thinspace cm}}

\def\erg{{\rm\thinspace erg}}

\def\K{{\rm\thinspace K}}
\def\keV{{\rm\thinspace keV}}
\def\km{{\rm\thinspace km}}

\def\Msun{\hbox{$\rm\thinspace M_{\odot}$}}

\def\s{{\rm\thinspace s}}

\def\ergps{\hbox{$\erg\s^{-1}\,$}}

\def\kmps{\hbox{$\km\s^{-1}\,$}}

\def\pcmsq{\hbox{$\cm^{-2}\,$}}

\slugcomment{ }

\shorttitle{Constraints on Compton-thick winds}
\shortauthors{C.~S.~Reynolds}

\begin{document}

\title{Constraints on Compton-thick winds from black hole accretion disks: can we see the inner disk?}

\author{Christopher~S.~Reynolds\altaffilmark{1,2}}
\email{chris@astro.umd.edu}
\altaffiltext{1}{Department of Astronomy, University of Maryland, College Park, MD 20742-2421, USA}
\altaffiltext{2}{Joint Space-Science Institute (JSI), College Park, MD 20742-2421, USA}

\begin{abstract}
Strong evidence is emerging that winds can be driven from the central regions of accretion disks in both active galactic nuclei (AGN) and Galactic black hole binaries (GBHBs).  Direct evidence for highly-ionized, Compton-thin inner-disk winds comes from observations of blueshifted ($v\sim 0.05-0.1c$) iron-K X-ray absorption lines.  However, it has been suggested that the inner regions of black hole accretion disks can also drive Compton-thick winds --- such winds would enshroud the inner disk, preventing us from seeing direct signatures of the accretion disk (i.e. the photospheric thermal emission, or the Doppler/gravitationally broadened iron K$\alpha$  line).      Here, we show that, provided the source is sub-Eddington, the well-established wind driving mechanisms fail to launch a Compton-thick wind from the inner disk.    For the accelerated region of the wind to be Compton-thick, the momentum carried in the wind must exceed the available photon momentum by a factor of at least $2/\lambda$, where $\lambda$ is the Eddington ratio of the source, ruling out radiative acceleration unless the source is very close to the Eddington limit.  Compton-thick winds also carry large mass-fluxes, and a consideration of the connections between the wind and the disk show this to be incompatible with magneto-centrifugal driving.  Finally, thermal driving of the wind is ruled out on the basis of the large Compton-radii that typify black hole systems.   In the absence of some new acceleration mechanism, we conclude that the inner regions of sub-Eddington accretion disks around black holes are indeed naked. 
\end{abstract}



\section{Introduction}\label{intro}

There is overwhelming evidence that active galactic nuclei (AGN) are powered by accretion onto a supermassive black hole via a rotationally supported accretion disk (Shakura \& Sunyaev 1973; Rees 1984).   The spectrum and variability of the electromagnetic radiation from these accretion disks is our principal diagnostic for studying the central black hole and the physics with which accretion disks extract gravitational potential energy of infalling matter.  Most type-1 AGN (i.e. those displaying broad optical/UV emission lines) possess a strong optical/UV/EUV bump, believed to be the thermal emission from the photosphere of the inner accretion disk ($r\lesssim 100r_g$ where $r_g=GM/c^2$ is the gravitational radius of the black hole with mass $M$).   Modeling of this ``big blue bump" can be used to determine the mass accretion rate and, together with the bolometric luminosity, the radiative efficiency of the disk (Davis \& Laor 2011).  In X-rays, 30--50\% of type-1 AGN (Nandra et al. 2007) show strongly broadened iron emission lines which, when interpreted as disklines, can be used to determine the physical condition of the inner disk ($r\lesssim 10r_g$), and even the spin of the black hole (Fabian et al. 1995; Reynolds \& Nowak 2003; Brenneman \& Reynolds 2006; Miller 2007).   Parallel studies have been conducted for the stellar-mass black holes; analyses of the thermal disk continuum (now in the soft X-ray band) validate the basics of the standard disk model (Shakura \& Sunyaev 1973; Novikov \& Thorne 1974; Gierli\'nski \& Done 2004), and black hole spin has been explored using both thermal continuum signatures and broadened iron lines (Miller et al. 2004; McClintock et al. 2007; Gou et al. 2010; Fabian et al. 2012).

All of these investigations operate under the assumption that we have a direct view of the photosphere of the inner accretion disk.    However, it is becoming clear that accretion disks drive winds that can impact the observables.    Blueshifted absorption lines from highly-ionized winds have been seen in radio-quiet AGN (Lee et al. 2001; McKernan, Yaqoob \& Reynolds 2007; Tombesi et al. 2010a), radio-loud AGN (Tombesi et al. 2010b; Torresi et al. 2012), and Galactic Black Hole Binaries (GBHBs; Miller et al. 2006; Neilsen \& Lee 2009; King et al. 2012).   In these cases, the wind is optically-thin (outside of the strongest resonant absorption lines).  Hence, the observed spectrum can be absorption-corrected and the inner-disk signatures recovered once the column density of each ionic species in the wind is determined. 

Compton-thick winds, on the other hand, affect the spectrum in a manner that all-but destroys our ability to reconstruct the inner-disk signatures.   In recent years, a number of authors (e.g. Sim et al. 2008, 2010a,b; Tatum et al. 2012) have discussed the possibility that a Compton-thick equatorial wind can be driven off the inner parts of the accretion disk ($r\sim 100r_g$) at the escape speed ($v_{\rm esc}\sim 0.1-0.2c$).     These models find that the wind can become strongly ionized and, when we view the central compact X-ray source through the wind, the spectrum becomes imprinted with deep and strongly blueshifted iron absorption lines/edges such as those found in the quasar PG1211$+$143 (Sim et al. 2010a).  However, even when we view the X-ray source along low-inclination lines-of-sight that do not intercept the wind, the reprocessing of the primary X-rays by a Compton-thick wind can produce significant complexity in the spectrum.   The ionized wind can be a source of iron line photons (through radiative recombination), and the iron line profile can develop a pronounced red-tail by Compton downscattering from the opposing side of the wind.  Indeed, Tatum et al. (2012) have successful described the broad iron lines of several unabsorbed (``bare") Seyfert galaxies as pure wind signatures.

In this {\it Letter}, we address the circumstances under which a Compton-thick wind can be generated by the inner regions of a black hole accretion disk.   By considering the wind's momentum- and mass-flux in the light of possible acceleration mechanisms, we show that the inner regions of sub-Eddington accretion disks (such as characterize many Seyfert nuclei as well as GBHBs in their thermal-dominant state) are incapable of generating such a Compton-thick wind.   This validates the methodology whereby observations are compared with ``naked" accretion disk models.    

The Letter is organized as follows.   Section~2 describes the wind model that we adopt and we write down expressions for the momentum and mass flux.  We then examine possible acceleration mechanisms.  After noting that pressure-driving cannot produce a wind from the inner accretion disk, we examine radiation- (Section~3) and magnetic-driving (Section~4).   The physics of these mechanisms impose constraints on the momentum- and mass-flux that are satisfied by confirmed wind systems but violated by the putative Compton-thick winds suggested to exist in some Seyfert nuclei.  We draw conclusions in Section~5.

\section{A simple wind model}

We shall adopt the simplest possible wind model.    Matter is injected into this wind (ultimately from the accretion disk) and accelerated radially such that, at radius $r=r_0$, it achieves terminal velocity $v_{\infty}$.  Thereafter the wind is assumed to have a constant velocity --- this is the {\it coasting zone} of the wind.   Following the notation of Knigge, Woods \& Drew (1995), we shall assume that $v_\infty=f_vv_{\rm esc}(r_0)$, where $v_{\rm esc}(r_0)=\sqrt{2GM/r_0}$ is the escape velocity from radius $r_0$.  The wind must have $f_v\gtrsim 1$ to avoid stalling and falling back to the disk.

If the density at the base of the coasting zone is $n_0$ then, given the assumptions above, mass continuity immediately implies that the density is $n(r)=n_0(r_0/r)^2$.   Assuming unity ionization fraction, the electron optical depth of the coasting zone is $\tau_e=\sigma_T\int_{r_0}^\infty n(r)\,dr=\sigma_Tn_0r_0$.   Hereafter, we use the term ``Compton-thick wind" to mean that the electron scattering optical depth of the coasting zone is greater than unity, $\tau_e>1$.

We suppose that this wind structure subtends a solid angle $\Omega$ as viewed from the center of the system.  With one exception, we scale the expressions in this paper to $\Omega=\pi$.    The precise geometry of the wind (e.g., equatorial vs. bipolar) is not important for most of the arguments in this paper.  When needed (Section~3), we assume an equatorial geometry.  

Let us define some useful quantities.  The mass flux in the wind is $\dot{M}_W=r^2\Omega nm_pv_{\infty}$.   We denote the (electromagnetic) bolometric luminosity of the inner disk as $L_{\rm bol}$, and the Eddington ratio as $\lambda=L_{\rm bol}/L_{\rm Edd}$ where $L_{\rm Edd}$ is the Eddington luminosity.  The mass accretion rate of the inner disk, $\dot{M}_A$, is related to $L_{\rm bol}$ via the radiative efficiency parameter $\eta=L_{\rm bol}/\dot{M}_Ac^2$.   We can also define an Eddington accretion rate by $\dot{M}_{\rm Edd}=L_{\rm Edd}/\eta c^2$.  Finally, for convenience, we define a scaled radius $x\equiv r_0/r_g$.

With these definitions, the mass flux of the wind is related to the optical depth by,
\begin{equation}
\dot{M}_W=\frac{1}{2\sqrt{2}}\,\tau_ef_vx^{1/2}\left(\frac{\Omega}{\pi}\right)\left(\frac{L_{\rm Edd}}{c^2}\right),
\end{equation}
giving a momentum-flux, $\dot{P}_W$,
\begin{equation}
\dot{P}_W=\dot{M}_Wv_\infty=\frac{2\,f_v^2}{\lambda}\left(\frac{\Omega}{4\pi}\right)\left(\frac{\tau_eL_{\rm bol}}{c}\right).
\label{eqn:pwtau}
\end{equation}
Another simple yet instructive form for the momentum flux is,
\begin{equation}
\dot{P}_W=\frac{\dot{M}_W}{\dot{M}_A}\dot{M}_Av_\infty=\sqrt{2}f_v\left(\frac{x}{100}\right)^{-1/2}\left(\frac{\eta}{0.1}\right)^{-1}\left(\frac{\dot{M}_W}{\dot{M}_A}\right)\frac{L_{\rm bol}}{c},
\label{eqn:pwmdot}
\end{equation}
which we can combine with eqn.~\ref{eqn:pwtau} to get,
\begin{equation}
\frac{\dot{M}_W}{\dot{M}_A}=\frac{1}{2\sqrt{2}}\frac{\tau_ef_v}{\lambda}\left(\frac{\Omega}{\pi}\right)\left(\frac{x}{100}\right)^{1/2}
\left(\frac{\eta}{0.1}\right).
\label{eqn:mdot_rat}
\end{equation}

We now proceed to use these expressions to examine wind acceleration mechanisms.  Adopting a magnetohydrodynamic (MHD) framework, there are three ways to accelerate a wind; radiative acceleration, pressure forces, or magnetic acceleration.  This statement follows immediately from the MHD force equation,
\begin{equation}
\rho\frac{D{\bf v}}{Dt}=-\nabla p+\frac{1}{4\pi}(\nabla\times {\bf B})\times {\bf B}+{\bf F}_{\rm rad}-{\bf g},
\end{equation}
(all symbols have their usual meaning, $D/Dt$ being the convective derivative) with the different acceleration mechanisms highlighting the relevance of the different force terms that can overcome gravity ${\bf g}$.   We will examine Compton-thick winds within the context of each of these mechanisms.

We begin by briefly addressing thermal driving of the wind which will be most powerful when the wind is at the Compton temperature of the radiation field, $T_C\sim 10^7-10^8$\,K for a typical AGN or thermal state GBHB.  The analytic theory of such winds was described by Begelman, McKee \& Shields (1983), and simulations of thermal winds were developed by Woods et al. (1996).   We define the Compton radius by $R_C=GMm_p/k_BT_C\approx 1.1\times 10^5T_8^{-1}r_g$ where $T_8\equiv T_C/10^8\,K$; this is the radius outside of which it is impossible to construct a hydrostatic equilibrium and an outflow must result.   In fact, due to disk rotation, a thermal wind can form at radii as small as $0.1R_C$.  However, for the range of Compton-temperatures under consideration, this still implies that thermal winds can only form on scales of $10^4r_g$ or more, with corresponding terminal velocities of $v\sim v_{\rm esc}(10^4r_g)\sim 4000\kmps$ or less.   While thermal driving is an excellent candidate for producing the observed warm absorbers seen in half of type-1 Seyfert nuclei (Reynolds 1997; George et al. 1998; McKernan et al. 2007), it cannot produce the inner disk winds that are our focus here.    These inner disk winds must be driven by radiation or magnetic forces.

\section{Radiative acceleration}

\begin{figure}[t]
\centerline{
\psfig{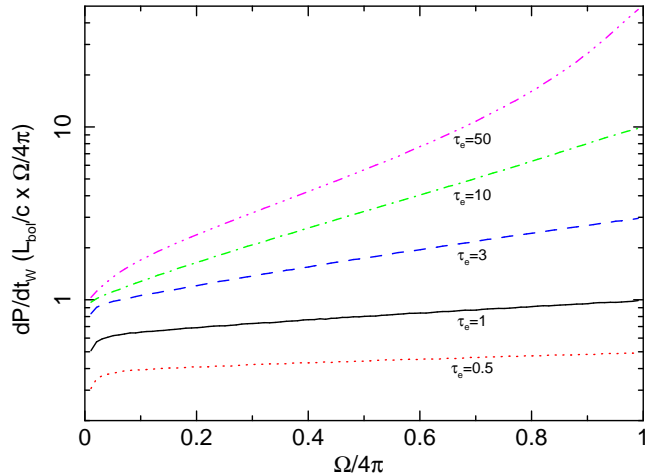}
}
\caption{{\small Momentum flux of a Compton-scattering-dominated equatorial wind accelerated by an isotropic radiation source as a function of the solid angle $\Omega$ subtended by the wind and the radial optical depth through the wind $\tau_e$.  The momentum flux is scaled to the (radial) momentum flux of the primary radiation field incident on the wind base ($=L_{bol}/c\times \Omega/4\pi$). }}
\label{fig:mc}
\end{figure}

Radiative acceleration occurs due to the transfer of momentum from the photon field to the wind.  In this discussion, we shall assume that the opacity of the wind is dominated by Compton scattering\footnote{For a plasma photoionized by a typical AGN or GBHB spectrum, this requires ionization parameters of $\xi\gtrsim 10^3\,{\rm erg}\,{\rm cm}\,{\rm s}^{-1}$ and is appropriate for the cases under consideration.}   If the wind has optical depth $\tau_e$ and completely surrounds an isotropic source of radiation with luminosity $L_{\rm bol}$, the momentum transfer to the wind is $\dot{P}_W=\tau_e L_{\rm bol}/c$.  For $\tau_e>1$, the multiple scattering of photons within the wind leads to a wind momentum (in some particular direction) that exceeds the photon momentum of the primary emission (in that direction).

In fact, there are two departures from spherical symmetry that affect the momentum transfer in Compton-thick disk winds.  Firstly, the wind may not enshroud the whole source.  Thus, the tangential optical depth may be appreciably less than the radial optical depth, allowing photons to escape the wind.   Secondly, by assumption, the mid-plane of the system is occupied by an optically-thick accretion disk, preventing scattered photons from traversing hemispheres.   To assess these effects, we specialize to an equatorial geometry, and perform a Monte Carlo calculation.   For a given $\Omega$ and $\tau_e$, we follow the Compton scattering of $2\times 10^5$ photons produced by a central isotropic source (with $r_{\rm emit}=0.01r_0$).   We use the (non-relativistic) angle-dependent Compton-scattering cross section and, in each scattering event, track the radial momentum transfer from the photons to the gas.   We do not track the change in the (black hole frame) photon energy resulting from the scattering events --- this will result in a slight {\it over-estimate} of the momentum transfer. A given photon is followed until it reaches a radius of $100r_0$ at which point we declare that it has escaped the wind.  If a photon is scattered into the disk-plane, we assume that it is absorbed and re-radiated isotropically.    We do, however, explicitly allow photons that leak into the evacuated bi-polar funnel to re-enter the wind if their trajectory should demand. Figure~\ref{fig:mc} shows the resulting momentum of the wind as a function of $\Omega$ and $\tau_e$.    For $\Omega/4\pi\ll1$, we find that photon leakage from the sides of the wind effectively eliminate multiple-scatterings and the wind momentum is $\dot{P}_W\approx\min(\tau_e,1)L_{\rm bol}/c\times (\Omega/4\pi)$.  We also see that, for very Compton-thick winds, even a small opening in the wind ($\Omega/4\pi=1-\epsilon, \epsilon\ll1$) can significantly decrease the momentum below the spherical-limit.

\begin{table*}[t]
\centerline{
\begin{tabular}{lcccccc}\hline\hline
Object & $L_{\rm bol}$  & $M_{\rm BH} $ & $\lambda$ & $\dot{M}_W/\dot{M}_A$& $\tau_e$& $\dot{P}_W$  \\
(Units) & ($10^{45}\ergps$) & $(10^7\Msun)$ & & & &  $(L_{\rm bol}/c\times \Omega/4\pi)$ \\\hline
Ark~120 	&		8.4	&	15	&	0.40	& 2.4	&	2.7	&13.4 \\
Fairall~9	&	11	&	26	&	0.32	& 1.7	&	1.6	&	9.8 \\
Mrk~335	&	1.8	&	1.4	&	0.92	& 0.56	&	1.4	&	3.14 \\
NGC7469	&	0.5	&	1.2	&	0.30	& 2.8	&	2.4	&	16.1 \\
SwiftJ2127.4 & 1.4	&	1.5	&	0.66	& 0.50	& 0.93 &	2.80\\\hline\hline
\end{tabular} 
}
\caption{Summary of wind parameters for the T12 sources, assuming $f_v=1$, $x=100$ and $\Omega=\pi$.   The bolometric luminosities are estimated using the absorption-corrected X-ray luminosities of T12 together with the bolometric correction factors of Vasudevan \& Fabian (2007).  For Ark~120, Fairall~9, Mrk~335, and NGC~7469, the bolometric luminosities are converted into Eddington ratios using black hole masses taken from the  reverberation mapping of Peterson et al. (2004).  For Swift~J2127.4$+$5654, we assume a black hole mass of $1.5\times 10^7\Msun$ based on the width of the H$\beta$ line (Malizia et al. 2008).  No good mass estimate exists for MCG--2-14-009 and hence we do not consider it further in this work.  The mass fluxes are calculated from the values of $\dot{M}_W/\dot{M}_{\rm Edd}$ quoted by T12.  }
\end{table*}

Thus, radiative-driving gives 
\begin{equation}
\dot{P}_W<\frac{\tau_eL_{\rm bol}}{c}\times \frac{\Omega}{4\pi}.
\label{eqn:radlim}
\end{equation}  
Combining this with eqn.~\ref{eqn:pwtau}, we see that radiation can drive a Compton-thick wind only if $\lambda>2f_v^2$, i.e., the source is radiating at (or above) the Eddington-limit and/or the wind is slower than the local escape speed.    Specifically, let us examine the Compton-thick winds inferred to exist in a number of ``bare" Seyfert nuclei by Tatum et al. (2012; hereafter T12).  They use a parameterized wind model coupled with a multi-dimensional radiative transfer code to fit the spectra of Ark~120, Fairall~9, MCG--2-14-009, Mrk~335, NGC~7469, and Swift~J2127.4$+$5654; in each case they can explain the broadened iron line as resulting from scattering in the wind and constrain the required mass flow rates.  In Table~1, we list the momentum fluxes and optical depths (from eqn.~\ref{eqn:mdot_rat}) corresponding to the T12 mass outflow rates (excluding MCG--2-14-009 due to the lack of a reliable black hole mass).   In all cases, we see that the momentum of the putative wind exceeds that expected from radiative driving.   

There is one subtlety that must be addressed.  Strictly, the optical depth that appears in the radiation-driving constraint (eqn.~\ref{eqn:radlim}) should include matter in the {\it accelerating zone} whereas our comparison with the T12 results only considers matter in the coasting zone. Does the presence of a high optical depth base to the wind rescue the radiative-driving scenario for these sources?   This question can be addressed using Figure~1.   For Mrk~335 and Swift~J2127.4, the momentum flux of the putative wind only exceeds that of the source photons by a factor of $\sim 3$.  A high optical depth ($\tau\sim 10$) base with a modest opening angle ($\Omega\sim 2\pi$) could, in principle, allow the wind to extract sufficient momentum from the photon field.  However, for Ark~120, Fairall~9, and NGC~7469, the momentum of putative wind exceeds that of the source photons by an order of magnitude.  Extracting the required momentum from the radiation field requires the wind base to possess a very high optical depth ($\tau>10$) and almost completely enshroud the central radiation source ($1-\Omega/4\pi\ll1$).  Compton downscattering in this cloud would produce an obvious spectral break in the observed X-ray continuum at $E_b\sim m_ec^2/\tau^2\lesssim 5\keV$.  The lack of any such break in the observed X-ray spectra of these AGN rules out this scenario.  

\section{Magnetic acceleration}

Finally, we examine winds driven by magnetic forces, specifically winds that are centrifugally accelerated down open, rotating, poloidal magnetic fields anchored in the disk (Blandford \& Payne 1982; also see review by Pudritz et al. 2007).   Since these winds are accelerated by the action of magnetic torques from fields that are embedded in the accretion disk, there is an intimate connection between the mass loss rate in the wind and the accretion onto the black hole.   The theory of MHD winds highlights two important {\it cylindrical} radii characterizing any given streamline; material is centrifugally lifted off the disk at the launch radius $R_L$ and continuously accelerated by magnetic forces until the flow becomes super-Alfv\'enic at radius $R_A>R_L$ (after which it coasts).  Hereafter, we denote the ratio of these radii by $\varpi\equiv R_A/R_L$; detailed models find that $\varpi\sim 2-3$ (Pudritz et al. 2007).   Conservation of a generalized Bernoulli parameter shows that $v_\infty=\varpi v_{\rm esc}(R_L)$. Most importantly, by considering the back reaction of the magnetic torques on the disk, it is found that 
\begin{equation}
\frac{\dot{M}_W}{\dot{M}_A}<\varpi^{-2},
\label{eqn:mhdmdot}
\end{equation}
(eqn.~15 of Pudritz et al. 2007).  Comparing Eqns.~\ref{eqn:mdot_rat} and \ref{eqn:mhdmdot}, we immediately see that a magnetocentrifugal wind will have difficulty in producing the mass fluxes required for Compton-thick winds.  To make the connection with our simple wind model more explicitly, we note that it is appropriate to relate $r_0$ to the end of the acceleration zone $R_A$ rather than the launch radius.  Thus, we have $v_\infty=\varpi v_{\rm esc}(R_L)=\varpi^{3/2} v_{\rm esc}(r_0)$, i.e., $f_v=\varpi^{3/2}$.  Substituting eqn.~\ref{eqn:mdot_rat} into eqn.~\ref{eqn:mhdmdot} we find that 
\begin{equation}
x<800\varpi^{-7}\frac{\lambda^2}{\tau_e^{2}}\left(\frac{\Omega}{\pi}\right)^{-2}\left(\frac{\eta}{0.1}\right)^{-2}.\label{eqn:mhd_constraint}
\end{equation}
Noting that $\varpi^{-7}\sim 2^{-7}\sim 10^{-2}$ we see that MHD torques will indeed fail to produce Compton-thick winds unless (i) the accretion rate is a significant fraction of the Eddington rate, (ii) the radiative efficiency is low, or (iii) the Alfv\'en radius is very close to the launch radius ($\varpi\approx 1$).   

\begin{figure}[t]
\centerline{
\psfig{figure=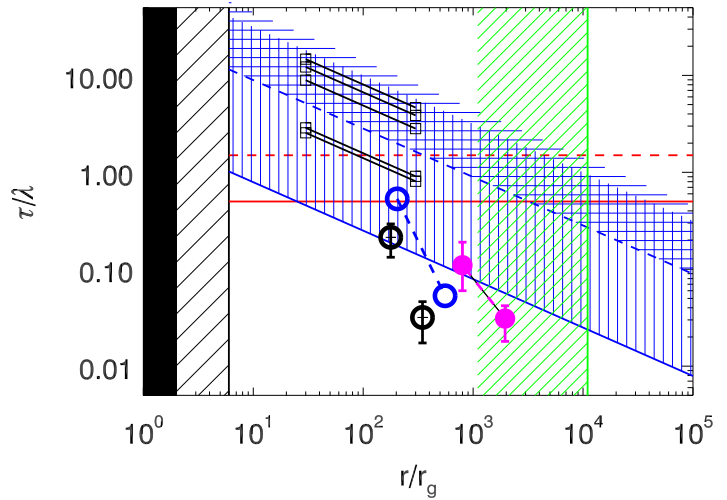,width=0.6\textwidth}
}
\caption{{\small Constraint diagram for wind-driving mechanisms, assuming $\eta=0.1$ and $\Omega=\pi$.   The solid black region lies within the event horizon.  The coarsely hashed region lies within the innermost stable circular orbit.   The blue (diagonal) lines shows the MHD driving constraint (eqn.~\ref{eqn:mhd_constraint}; region above forbidden) for $\varpi=2$ (solid) and $\varpi=1$ (dashed).  The green (vertical) line shows the thermal driving constraint $r>0.1R_C$ assuming a Compton temperature of $10^8\K$ (left region forbidden).   The red (horizontal) lines mark $\dot{P}_W=L_{\rm bol}/c\times \Omega/4\pi$ (solid) and $\dot{P}_W=3L_{\rm bol}/c\times \Omega/4\pi$ (dashed), thereby indicating the viability of radiative acceleration.   Also shown are wind parameters for IGR~J17091--3624 (magenta filled circles), PG1211$+$143 (blue open circles), the UFOs in 3C111/3C120 (black open circles), and the putative Compton-thick winds in bare Seyfert nuclei (diagonal black lines).}}
\label{fig:constraint}
\end{figure}

Figure~\ref{fig:constraint} (blue diagonal lines) displays the MHD-driving constraint on the $(\tau/\lambda,r)$-plane.  Taking the extreme case of $\varpi=1$ (dashed line), we find that MHD-driving becomes viable for the putative winds in the two near-Eddington sources of T12, Mrk~335 and Swift~J2127.4.  Even in this extreme limit, however, MHD-driving fails to drive Compton-thick winds in Ark~120, Fairall~9 and NGC~7469.

Having discussed winds that have been indirectly inferred, it is instructive to test our constraint diagram (Fig.~\ref{fig:constraint}) against  some of the more extreme {\it confirmed} wind systems. King et al. (2012, hereafter K12) reports a fast ($v=0.032c$) highly-ionized ($\log\xi\approx 3.3$) wind in the outbursting GBHB system IGR~J17091--3624.   The signatures of the wind are iron-K absorption lines in the {\it Chandra} High-Energy Transmission Grating (HETG) spectrum, and photoionization modeling suggests a column density of $N=4.7^{+1.7}_{-1.9}\times 10^{21}\pcmsq$ ($\tau_e=3.1^{+1.1}_{-1.3}\times 10^{-3}$).   K12 also report the tentative detection of a more highly ionized ($\log\xi\approx 3.9$), faster ($v=0.05c$) system with a column of $1.7^{+1.2}_{-0.8}\times 10^{22}\pcmsq$ ($\tau_e=1.1^{+0.8}_{-0.5}\times 10^{-2}$).   We put these two systems onto Fig.~\ref{fig:constraint} (magenta circles), assuming $v=v_{\rm esc}(r)$ to get the distance and taking $\lambda=0.1$ on the basis of the fact that the system was in outburst when the winds were observed.  Supporting the hypothesis of K12, we find that both of these systems can readily be explained as MHD-driven.  

One of the most extreme confirmed and persistent AGN winds occurs in the luminous ($\lambda=1$) Seyfert galaxy PG1211$+$143 (Pounds et al. 2003; Pounds \& Reeves 2009).  A detailed analysis of iron K-shell absorption lines in the {\it XMM-Newton} data for this object reveals two absorption systems with outflow velocities $v_1=0.06c$ and $v_2=0.1c$, and column densities $N_1=8\times 10^{22}\pcmsq$ and $N_2=8\times 10^{23}\pcmsq$ ($\tau_1=0.05$ and $\tau_2=0.5$).  From Fig.~\ref{fig:constraint}, we see that the slower system can readily be explained as either an MHD-driven or radiatively-driven wind.  The faster, higher-column system requires $\varpi\lesssim 1.5$ for MHD-driving to be viable.   However, the fact that this system has $\dot{P}_W\approx L_{\rm bol}/c\times \Omega/4\pi$ strongly implicates radiative driving.

Transient ultra-fast outflows (UFOs) have been found in a number of AGN by Tombesi et al. (2010a,b).   Here, we focus on the high-quality characterizations of the UFOs in two broad line radio galaxies, 3C111 and 3C120, using {\it Suzaku} observations of iron-K absorption lines.   For 3C111, Tombesi et al. (2011) report a system with $v=0.11c$ and column density $N=(7.7\pm 2.9)\times 10^{22}\pcmsq$.  Using the black hole mass of $M=2.4\times 10^8\Msun$ (Chatterjee et al. 2011) and a bolometric luminosity of $L_{\rm bol}\sim 8\times 10^{45}\ergps$ (Tombesi et al. 2011), we estimate an Eddington ratio of $\lambda_{3C111}=0.24$.   For 3C120, which has a very similar Eddington ratio of $\lambda_{3C120}\approx 0.23$ (Cowperthwaite \& Reynolds 2012), Tombesi et al. (2010b) find a wind system with $v=0.076c$ and $N=(1.1\pm 0.5)\times 10^{22}\pcmsq$.  The constraint diagram reveals that these UFOs can readily be driven by magnetic forces. 

\section{Conclusions}

We have used mass- and momentum-flux arguments to examine the driving mechanisms of winds from the inner regions of black hole accretion disks.  We find that the fast, optically-thin winds discovered in both GBHBs and AGN via iron K-shell absorption lines can be readily explained as either magnetically-driven or (in the case of the Eddington-limited source PG1211$+$143) radiation-driven.  Reasonable extensions of the radiative- and/or MHD-driving pictures can also explain the winds that are inferred indirectly to exist in the near-Eddington sources Mrk~335 and Swift~J2127.4 (T12).   However, we fail to identify a viable driving mechanism for the putative Compton-thick winds in the sub-Eddington systems of T12.  In the absence of some new driving mechanism, we conclude that these systems do not possess Compton-thick winds and that the observed spectral complexity originates from the accretion disk itself, lines being broadened through the usual combination of Doppler and gravitational redshifts.



\acknowledgments
\section*{Acknowledgments}
The author thanks Andy Fabian, Jon Miller and Ralph Pudritz for insightful discussions as well as the anonymous referee for helpful comments that significantly improved the discussion of radiative acceleration.  Much of this work was performed during a visit to the Max Planck Institut f\"ur Astrophysik, Garching, and the author thanks the Directors of MPA for their hospitality.  The author thanks support from NASA (Suzaku GO grant NNX10AR31G and ADAP grant NNX12AE13G).

\clearpage

\end{document}